\documentclass[preprint,final,3p,times,12pt,authoryear]{article}
\usepackage{graphicx} 
\usepackage{amsmath,amssymb,amstext}
\usepackage{bm}
\usepackage{gensymb}
\usepackage{siunitx}
\usepackage{multirow}
\usepackage{caption}
\usepackage{subcaption}
\usepackage{algorithm}
\usepackage{algpseudocode}
\usepackage{multirow}
\usepackage{placeins}
\usepackage{pdflscape}
\usepackage{geometry}
\usepackage{hyperref}
\usepackage{natbib}

\begin{document}

\title{Spatio-temporal data fusion for the analysis of in situ and remote sensing data using the INLA-SPDE approach}

  \author{Shiyu He and Samuel W.K. Wong\thanks{Author for correspondence: samuel.wong@uwaterloo.ca}\hspace{.2cm}\\
	Department of Statistics and Actuarial Science, University of Waterloo}
\maketitle

\begin{abstract}
We propose a Bayesian hierarchical model to address the challenge of spatial misalignment in spatio-temporal data obtained from in situ and satellite sources. The model is fit using the INLA-SPDE approach, which provides efficient computation. Our methodology combines the different data sources in a ``fusion'' model via the construction of projection matrices in both spatial and temporal domains. Through simulation studies, we demonstrate that the fusion model has superior performance in prediction accuracy across space and time compared to standalone ``in situ" and ``satellite" models based on only in situ or satellite data, respectively. The fusion model also generally outperforms the standalone models in terms of parameter inference. Such a modeling approach is motivated by environmental problems, and our specific focus is on the analysis and prediction of harmful algae bloom (HAB) events, where the convention is to conduct separate analyses based on either in situ samples or satellite images. A real data analysis shows that the proposed model is a necessary step towards a unified characterization of bloom dynamics and identifying the key drivers of HAB events.
\end{abstract}

\noindent
{\it Keywords:} Data fusion, Spatial misalignment, Integrated Nested Laplace Approximation (INLA), Stochastic partial differential equations (SPDE)

\section{Introduction}

Spatial misalignment has become a challenge along with the technological advances in the collection and storage of spatio-temporal data, such as the use of GPS, satellites, etc. 
While convenient for undertaking large-scale spatio-temporal analyses, it can be difficult to integrate or compare data derived from different sources, when each has its own spatial and temporal reference frames. These data might be obtained from a network of monitoring stations, satellites, or numerical outputs of computer simulations. While data collected from monitoring networks give measurements at points in space, remote sensing observations and outputs from computer simulations provide predictions at the level of grid cells \citep{bruno2016survey}. In spatial statistics, the challenge brought by data measured at different scales is commonly known as the change of support problem (COSP) \citep{cressie2015statistics, banerjee2003hierarchical, gotway2002combining}. Data from monitoring networks are classified as point-referenced or geostatistical data, and often tend to be sparsely located in both spatial and temporal dimensions. When available, these data are generally considered to accurately represent the true conditions or phenomena being measured \citep{berrocal2010spatio}. On the other hand, measurements from satellites provide high-resolution data on grid cells at regular time intervals, are readily accessible, and it is desirable to leverage such massive data for analyses. However, care must be taken to ensure seamless compatibility across different data sources and to ensure the reliability of analyses. The associated methodological challenges are particularly significant in fields that include epidemiology, urban planning, and environmental science.

In the realm of environmental science, the integration of data from sources such as satellites, monitoring stations, and field measurements is pivotal for the understanding of ecological changes and climatic trends. In the spatio-temporal modeling of harmful algae bloom (HAB) dynamics, analyses tend to rely heavily on either in situ samples or satellite images in isolation. Within this field, common practices involve constructing an algae bloom index followed by categorization as ``in situ" or ``remote sensing" \citep{fang2019space, manning2019extending, stumpf2016forecasting}. However, these practices have contributed to disparities in the characterization of bloom dynamics, leading to divergent findings regarding the main drivers of HABs \citep{ho2015challenges}. Indeed, integrating different types of observations was recognized as a key research area to progress towards a better predictive knowledge of HAB dynamics \citep{bullerjahn2016global}. Spatial misalignment also occurs in epidemiology studies for understanding the relationship between environmental factors and human health \citep{fuentes2006spatial, zhu2003hierarchical}, as well as environmental prediction and model calibration \citep{lawson2012bayesian}. Thus, to ensure reliable and accurate analyses in such application areas, spatial misalignment needs to be addressed in a modeling framework.

Different methods have been proposed to address the spatial misalignment problem, including block kriging \citep{banerjee2003hierarchical, cressie2015statistics, gotway2007geostatistical}, hierarchical downscaler \citep{berrocal2010spatio}, and Bayesian melding \citep{fuentes2005model}. 
The utilization of Bayesian models, particularly those developed through latent point-level processes has garnered significant attention, as demonstrated by the work of \cite{fuentes2005model}. This approach assumes that there exists an underlying unobserved spatial process driving both the observational data and the numerical model output, and can be implemented via Markov chain Monte Carlo (MCMC) methods. 
However, this type of state space framework requires stochastic integration of the latent spatial process over the area of interest, which is a computationally demanding task. To mitigate the computational bottleneck, fast and approximate methods, including the integrated nested Laplace approximation (INLA) \citep{rue2009approximate} and stochastic partial differential equation (SPDE) \citep{lindgren2011explicit} approaches, have been proposed. To address COSP, \cite{moraga2017geostatistical} proposed a method for the combined analysis of point and areal data utilizing the INLA-SPDE framework.

Spatio-temporal data fusion is an important research area in environmental science; however, it is much more challenging to tackle due to the higher dimensionality introduced by time. Following \cite{fuentes2005model}'s Bayesian melding approach, a spatio-temporal fusion model has been proposed to study the association between mortality and
pollution exposure to daily fine particulate matter, and addresses the COSP between point-referenced data and model output \citep{choi2009spatial}. Further, \cite{mcmillan2010combining} presents a spatio-temporal fusion model at the grid-cell level by assuming that the data process at the point level is linked to the same underlying true process at the grid-cell level. These models rely on MCMC methods for inference, so do not scale well to much higher dimensions due to the computational requirements of MCMC. 
While the INLA-SPDE-based approach has been proposed to model spatio-temporal processes \citep{blangiardo2015spatial, cameletti2013spatio}, its application to spatio-temporal data fusion is limited. \cite{villejo2023data} proposed a two-stage estimation approach to address spatial misalignment in linking pollutant exposures and health outcomes; however, the COSP was not specifically addressed in their first stage model for the pollutants. In their work, satellite data were not treated as block data; instead, they used the centroid of the image pixels as point data. Thus, to the best of our knowledge, there has been no previously proposed spatio-temporal data fusion approach that explicitly addresses the COSP through the INLA-SPDE framework.

The motivation of this paper is to address the COSP between different data sources, with application to better quantifying the disparities in characterizing the dynamics of HABs. To achieve this goal, we propose a Bayesian hierarchical model to analyze spatially misaligned data with a latent spatio-temporal dynamic process. The model considers the COSP and coherently combines data with different spatio-temporal resolutions and potential missingness. We show that via the construction of appropriate projection matrices, the different types of data can be spatially aligned and fitted using the INLA-SPDE approach in a computationally efficient manner. A real data analysis of HABs is presented, and the results illustrate the necessity of the fusion model for accurately characterizing their spatio-temporal dynamics. 

The structure of the paper is presented as follows. In Section \ref{sec:method}, we describe our proposed spatio-temporal fusion model to address the spatially misaligned data. Section \ref{sec:inference} discusses model inference and the application of INLA and the SPDE approach for  fitting the proposed model. Section \ref{sec:simulation} provides a simulation study inspired by the cyanobacteria HAB data to evaluate model performance. In Section \ref{sec:application}, we present a real-world example analyzing cyanobacteria HAB data over Western Lake Erie. Finally, Section \ref{sec:discussion} concludes the paper and outlines directions for future work.

\section{Method}\label{sec:method}

\subsection{Spatial data fusion model}
We first describe the spatial data fusion framework known as the Bayesian melding model; such a model structure can be found in \cite{fuentes2005model}. The model is designed to integrate both point and areal-level data by assuming a common spatial random field $\{y(\bm{s}):\bm{s} \in D \subset \mathbb{R}^2\}$ underlying all observations on the spatial domain of interest $D$, with mean $E(y(\bm{s})) = 0$ and a stationary and isotropic Mat\'ern covariance function. This spatial process is, in part, explained by large-scale explanatory variables $\bm{x}(\bm{s}) = (\bm{x}_0(\bm{s}), \dots, \bm{x}_p(\bm{s}))^T$ (i.e., a vector of $p$ covariates and an intercept), with $\bm{\beta} = (\beta_0, \dots, \beta_p)^T \in \mathbb{R}^{p+1}$ being the vector of coefficients:
\begin{equation}
   y(\bm{s}) = \bm x(\bm{s}) \bm\beta + \xi(\bm{s}),
\end{equation}
where the residual component $\xi(\bm{s})$ is modeled as a Gaussian random field.

On the observation level, denote the remote sensing observations as $z_1(\bm{B}_j)$ for block area $\bm{B}_j \in D \subset \mathbb{R}^2, j = 1,\dots, J$, and denote the in situ observations as $z_2(\bm{s}_i)$ for $\bm{s}_i \in D \subset \mathbb{R}^2, i = 1, \dots, I$. More generally, $z_1(\bm{B}_j)$ can be extended to model any areal data, and $z_2(\bm{s}_i)$ can be extended to model any point-referenced data. The two sets of observations are assumed to be generated with the following framework:
\begin{align}
    z_1(\bm{B}_j) & = a(\bm{B}_j) + \frac{1}{|\bm{B}_j|}\int_{\bm{B}_j} y(\bm{s}) d\bm{s} + e_1(\bm{B}_j), \\
    z_2(\bm{s}_i) & = y(\bm{s}_i) + e_2(\bm{s}_i),
\end{align}
where $e_1(\bm{B}_j) \sim N(0, \tau_1)$ and $e_2(\bm{s}_i) \sim N(0, \tau_2)$ are all independent. The parameter $a(\bm{B}_j)$ is the bias of the satellite images, $|\bm{B}_j|$ represents the area of block $j$ and hence $\frac{1}{|\bm{B}_j|}\int_{\bm{B}_j} y(\bm{s}) d\bm{s}$ represents the block average of the spatial random process within block $j$. Since satellite data tend to be more prone to noise and less associated with the true latent field compared to in situ data collected directly from ground monitoring stations, the introduced bias parameters allow the model to properly calibrate for such a discrepancy when present. Bias parameters are assumed to be either constant or spatially varying, although they are typically treated as constant across space to avoid identifiability issues. Note that this model set up allows for addressing the COSP, as satellite data are represented as block averages. \cite{moraga2017geostatistical} in their paper reformulate the above structure using $\bm \mu(\bm{s}) = \bm x(\bm{s}) \bm\beta$, such that $z_2(\bm{s}_i) = \mu(\bm{s}_i) + \bm \xi(\bm{s}_i) + e_2(\bm{s}_i)$, and $z_1(\bm{B}_j) = \frac{1}{|\bm{B}_j|}\int_{\bm{B}_j} (a + \mu(\bm{s}) + \bm\xi(\bm s)) d\bm{s} + e_1(\bm{B}_j) = \frac{1}{|\bm{B}_j|}\int_{\bm{B}_j} (\mu^*(\bm{s}) + \bm\xi(\bm s)) d\bm{s} + e_1(\bm{B}_j)$ where $\mu^*(\bm{s}) = a + \mu(\bm{s})$.

\subsection{Spatio-temporal data fusion model}

We propose an extension of the spatial data fusion framework to the spatio-temporal domain. The spatio-temporal extension to the data fusion framework is designed to borrow information and conduct inference across both space and time. The model configuration is explained as follows for completely-observed data. The remote sensing observations $z_1(\bm{B}_j, t)$ for each block area $\bm{B}_j \in D \subset \mathbb{R}^2, j = 1, \dots, J,$ at discrete time $t, t = 1, \dots, T$, are modeled as 
\begin{equation}\label{eqn:satellite}
    z_1(\bm{B}_j, t) = a(\bm{B}_j) + \frac{1}{|\bm{B}_j|}\int_{\bm{B}_j} y (\bm{s}, t) d\bm{s} + e_1(\bm{B}_j, t),
\end{equation}
where $a(\bm{B}_j)$ is the bias term assumed to be constant across time, $y(\bm{s}, t)$ is the hidden spatio-temporal process that varies continuously over space and time, and $e_1(\bm{B}_j, t) \sim N(0, \tau_1)$ is the measurement error for remote sensing. On the other hand, the in situ data $z_2(\bm{s}, t)$ are modeled as
\begin{equation}\label{eqn:insitu}
    z_2(\bm{s}_i, t) = y (\bm{s}_i, t) + e_2(\bm{s}_i, t),
\end{equation}
for $\bm{s}_i \in D \subset \mathbb{R}^2, i = 1, \dots, I, t = 1, \dots, T$, where $e_2(\bm{s}_i, t) \sim N(0, \tau_2)$ is the measurement error for the in situ data. 

Similar to the spatial setup, the spatio-temporal process is assumed to be a Gaussian random field underlying both sets of observations. The hidden process is inspired by models developed in air quality literature for modeling pollutants \citep{cameletti2011comparing, cameletti2013spatio, cocchi2007hierarchical}, which exhibits flexibility in capturing various factors influencing spatio-temporal dynamics. The hidden spatio-temporal process $y(\bm{s}, t)$ is assigned the following structure:
\begin{equation}
    y (\bm{s}, t) = \bm x(\bm{s}, t) \bm\beta + \xi(\bm{s},t),\label{eqn:spatio-temporal-covariate}
\end{equation}
where $\bm x(\bm{s}, t)$ is the set of large-scale spatially and temporally varying covariates, such as meteorological and geographical variables, although some of these may only vary on the spatial domain and others on the temporal domain. The latent process $\xi(\bm{s}, t)$ accounts for spatial and temporal dependencies and is assigned an AR(1) (i.e., first-order autoregressive) Gaussian process:
\begin{equation}
     \xi(\bm{s},t) = \rho \xi(\bm{s}, t-1) + \omega(\bm{s}, t), \label{eqn:spatio-temporal}
\end{equation}
for $t=2, \dots, T$, which is stationary with $|\rho| <1$, and $\xi(\bm{s}, 1) \sim N(0, \sigma_\omega^2/(1-\rho^2))$.
Further, $\omega(\bm{s}, t)$ is a spatially correlated innovation term, following a Gaussian distribution with a Mat\'ern covariance structure that is independent of time:
\begin{equation}\label{eqn:covfun}
    Cov(\omega(\bm{s}_i, t), \omega(\bm{s}_j, t')) = 
    \begin{cases}
        0, & \quad t \neq t',\\
        \frac{\sigma_\omega^2}{\Gamma(\nu)2^{\nu-1}}(\kappa||\bm{s}_i - \bm{s}_j||)^{\nu}K_{\nu}(\kappa||\bm{s}_i - \bm{s}_j||), & \quad t = t'.
    \end{cases}
\end{equation}
The above specification indicates when $t \neq t'$, the innovation term has zero covariance, i.e., $\omega(\cdot, \cdot)$ does not contribute to the spatial dependence between locations for measurements taken at different times. When $t = t'$, the covariance has a Mat\'ern structure that depends on $||\bm{s}_i - \bm{s}_j||$, i.e., the Euclidean distance between two generic locations $\bm{s}_i, \bm{s}_j \in \mathbb{R}^2$. The covariance function assumes the Gaussian field is second-order stationary and isotropic between locations. The term $K_\nu$ denotes the modified Bessel function of the second kind with order $\nu > 0$, which is usually kept fixed and measures the degree of the smoothness of the process. The parameter $\sigma_\omega^2$ is the marginal variance, and $\kappa > 0$ is a scaling parameter associated with the range $r$, i.e. the distance at which the spatial correlation becomes very small. Typically, the empirically derived definition for the range is $r = {\sqrt{8\nu}}/{\kappa}$,
with $r$ corresponding to the distance at which the spatial correlation is close to 0.1, for each $\nu \geq 0.5$.

\section{Inference}\label{sec:inference}

\subsection{INLA-SPDE}
The INLA method, proposed by \cite{rue2009approximate}, provides a computationally efficient approach for estimating posterior distributions in a broad class of latent Gaussian models. Unlike MCMC, INLA employs a deterministic approximation technique, including Laplacian approximations and numerical integration techniques, that allows for fast approximate Bayesian inference. INLA-SPDE extends the capabilities of INLA by drawing the link between the Gaussian field (GF) and Gaussian Markov random field (GMRF), and has been shown to be effective for handling complex spatial processes \citep{lindgren2011explicit}.  

In equation \eqref{eqn:spatio-temporal}, the latent process $\omega(\bm{s}, t)$ with a Mat\'ern covariance of form \eqref{eqn:covfun} is a solution to a certain SPDE; this SPDE can be represented as a discretely indexed GMRF by means of a finite basis function defined on a triangulation of the region of study. The finite element representation of the SPDE establishes a link between the GF and the GMRF characterized by the Gaussian weights with Markov dependencies.
The SPDE that can be used to construct the GMRF representation is
\begin{align}\label{eqn:pde}
    (\kappa^2 - \Delta)^{\alpha/2}(\tau_\omega\omega(\bm{s}, t)) = \mathcal{W}(\bm{s}, t), \quad \alpha = \nu + d/2, \quad \kappa >0, \quad \nu >0,
\end{align}
where $d=2$ for $\bm{s} \in \mathbb{R}^2$, $(\kappa^2 - \Delta)^{\alpha/2}$ is a pseudo-differential operator, and $\Delta = \sum_{i=1}^{d}\frac{\partial^2}{\partial \xi_i^2}$ is the Laplacian operator.  The parameter $\kappa > 0$ controls the scale, and $\tau_\omega$ controls the variance of the GF respectively. The parameter $\alpha$ is the order of the differential operator and controls the smoothness of the GF. The marginal variance $\sigma_\omega^2$ of the GF can be derived as $\sigma_\omega^2 = 1/{(4\pi) \kappa^{2}\tau_\omega^2}$.
In addition, the innovation term $\mathcal{W}(\bm{s},t)$ is the Gaussian spatial white noise process with unit variance. 

The solution to the SPDE in \eqref{eqn:pde} is a finite element representation with piecewise linear basis functions defined on a triangulation of $D$, which is given by
\begin{align}
    \omega(\bm s, t) = \sum_{g=1}^{G}\varphi_g(\bm s)\tilde\omega_{gt},
\end{align}
where $G$ is the total number of vertices of the triangulation, $\{\varphi_g(\cdot)\}_{g=1}^{G}$ is the set of basis functions, and $\{\tilde\omega_{gt}\}_{g=1, \dots, G, t=1,\dots,T}$ are zero mean Gaussian distributed weights. The weights $\tilde\omega_{gt}$ are updated over time in tandem with the temporal evolution of the GMRF, however, the set of basis functions remains constant over time, as the mesh associated with the triangulation remains fixed. 

Consequently, $\bm{\omega}_t := \bm\omega(\cdot, t)$ for each $t$ is a GMRF with distribution $N(0, \bm Q_{S}^{-1})$ and represents the approximated solution to the SPDE. Denote $\bm{\xi}_t := \bm{\xi}(\cdot, t)$ and $\bm{\xi} := (\bm{\xi}_1, \dots, \bm{\xi}_T)^T$, then \eqref{eqn:spatio-temporal} can be rewritten as $\bm{\xi}_t = \rho\bm{\xi}_{t-1} + \bm{\omega}_t$. Denote $\bm{y}_t := \bm{y}(\cdot, t)$ and $\bm{x}_t := \bm{x}(\cdot, t)$, then \eqref{eqn:spatio-temporal-covariate} can be rewritten as $\bm{y}_t = \bm{x}_t\bm{\beta} + \bm{B}\bm{\xi}_t$, where $\bm{B}$ is a sparse matrix that connects $\bm{\xi}_t$ to each observational vector $\bm{y}_t$. Therefore, the distribution of $\bm{\xi}|\bm{\Theta}$ is
\begin{equation}
    \bm{\xi}|\bm{\Theta} \sim N(0, \bm{Q}^{-1}),
\end{equation}
where $\bm{Q} = \bm{Q}_T \otimes \bm{Q}_S$, and $\bm{Q}_T$ is the precision matrix of the AR(1) process. The details of $\bm Q_{S}$ and $\bm Q_{T}$ can be found in \cite{lindgren2011explicit} and \cite{cameletti2013spatio}. The latent GF of the model is therefore $\bm{\mathcal{X}} := (\bm\xi, a, \bm\beta)$ and the hyper-parameters are $\bm{\Theta} := (\kappa, \tau_\omega, \rho, \tau_1, \tau_2)$. Denote $\bm{z}_{1t} := \bm{z}_1(\cdot, t)$ and $\bm{z}_{2t} := \bm{z}_2(\cdot, t)$, the joint posterior of $\bm{\mathcal{X}}$ and $\bm{\Theta}$, given complete observations $\bm{z}$, can thus be written as
\begin{equation}
    p(\bm{\mathcal{X}}, \bm{\Theta}|\bm{z}) = p(\bm{\Theta})p(\bm{\mathcal{X}}|\bm{\Theta})\prod_{t=1}^{T}p(\bm{z}_{1t}|\bm{\mathcal{X}}, \bm{\Theta})\prod_{t=1}^{T} p(\bm{z}_{2t}|\bm{\mathcal{X}}, \bm{\Theta}), \label{eq:inlaposterior}
\end{equation}
where $\bm{z}_{1t}|\bm{\mathcal{X}}, \bm{\Theta} \sim N(a + \frac{1}{|\bm{B}_j|}\int_{\bm{B}_j} (\bm{x}_t \bm{\beta} + \bm{B}\bm{\xi}_t) d\bm{s}, \tau^{-1}_1\bm{I})$, $\bm{z}_{2t}|\bm{\mathcal{X}}, \bm{\Theta} \sim N(\bm{x}_t \bm{\beta} + \bm{B}\bm{\xi}_t, \tau^{-1}_2\bm{I})$, and $p(\bm{\Theta})$ is the prior distribution of the hyper-parameters.
Finally, given this construction of $p(\bm{\mathcal{X}}, \bm{\Theta}|\bm{z})$, the marginal posterior distributions of $p(\bm{\mathcal{X}}|\bm{z})$ and $p(\bm{\Theta}|\bm{z})$ can be approximated through INLA.

\subsection{Missing values in response}

In the context of satellite observations, missing values in the response (i.e., unobserved blocks) are prevalent within the daily satellite images due to factors such as the presence of cloud cover. Denote the complete satellite data as $\bm{z}_2 = (\bm{z}_{2,obs}, \bm{z}_{2,mis})$, where $\bm{z}_{2,obs}$ and $\bm{z}_{2,mis}$ are the observed and missing components of $\bm{z}_2$, respectively. Since the missingness mechanism depends on factors apart from the response itself, we may treat the missingness mechanism of the satellite data as ignorable, i.e., it does not affect the specification of the posterior distribution \citep{little2019statistical}. Specifically, the term $p(\bm{z}_{2t}|\bm{\mathcal{X}}, \bm{\Theta})$ in the posterior distribution \eqref{eq:inlaposterior} may have missing responses for a different subset of blocks at each time $t$. INLA handles these missing values by computing their predictive distribution \citep{gomez2020bayesian}; in this context, the predictive distribution of an individual missing response $z_{mis}$ in the satellite data is
\begin{equation}
    p(z_{mis}|\bm{z}_1, \bm{z}_{2,obs}) = \int p(z_{mis}, \bm{\Theta} |\bm{z}_1, \bm{z}_{2,obs}) d\bm{\Theta} = \int p(z_{mis} |\bm{\Theta}, \bm{z}_1,\bm{z}_{2,obs}) p(\bm{\Theta} |\bm{z}_1,\bm{z}_{2,obs}) d\bm{\Theta}.
\end{equation}

\subsection{Approximation of integrals}
In the spatial fusion method using INLA-SPDE \citep{moraga2017geostatistical}, the integral of the spatial process in the areal data is approximated through:
\begin{equation}
    \int_{B_j} y(\bm{s})d\bm{s} \approx \sum_{g=1}^{G} A^a_{jg}\varphi_g = \bm{A^a_{j\cdot}\varphi},
\end{equation}
where $G$ is the number of vertices in the triangulation, $\bm{\varphi}$ is the vector of finite element basis functions, and $\bm{A^a} \in \mathbb{R}^{J \times G}$ is a projection matrix that maps the GMRF from the $J$ areal observation locations to the $G$ triangulation vertices. 

We extend such an approximation to the spatio-temporal integration, with the integral of the areal data being approximated as:
\begin{equation}
    \int_{B_j} y(\bm{s}, t)d\bm{s} \approx \sum_{g=1}^{G} \bm{[A^a_t]}_{jg}\varphi_{g} = \bm{[A^a_t]_{j\cdot}\varphi},
\end{equation}
where for each $t, t=1,\dots, T$, there exists a projection matrix $\bm{A^a_t} \in \mathbb{R}^{J \times G}$ that maps from the spatio-temporal GMRF to the areal observations. 
Each weight matrix $\bm{A^a_t}$ is a sparse matrix with row-wise entries summing to $1$. For vertex $g$ outside the area $j$, the entry $[\bm{A_t^a}]_{jg} = 0$. For vertices $g_1, \dots, g_m$ inside the area $j$,  each of them is assigned an equal weight of $1/m$ for mapping from the areal region to each triangulation vertex, i.e., $[\bm{A_t^a}]_{g_m,j} = 1/m$.  Since the triangulation vertices at time $t$ are only relevant for mapping GMRF at time $t$, $\bm{A^a}$ for the spatio-temporal integral approximation is a sparse block diagonal matrix of the form 
\begin{equation}\label{eqn:mapping_a}
    \bm{A^a} = \begin{bmatrix}
        \bm{A^a_{1}} & 0 & \cdots & 0 \\
        0 & \bm{A^a_{2}} & & \vdots \\
        \vdots &  & \ddots & 0 \\
         0 & \cdots & 0 & \bm{A^a_{T}}
    \end{bmatrix}.
\end{equation}
The construction of this projection matrix enables the fitting of the fusion model using the INLA-SPDE approach, thereby bypassing the need for expensive MCMC computations. Note that the mesh size plays a crucial role in defining the precision of the spatial field. A finer mesh provides a more accurate representation of the spatial domain but comes at the cost of increased computational complexity. On the other hand, a coarser mesh reduces computational demands but may sacrifice accuracy in representing spatial features. The practical effects of mesh size are explored in the subsequent simulation study.

We use non-informative/weakly informative priors for both fixed effects and the hyper-parameters. The priors for fixed effects $a$ and $\bm\beta$ are chosen to be normal with mean zero and large variance. The prior for the precision hyper-parameters are $\log \tau_1, \log \tau_2 \sim \log \Gamma (0.01, 0.01)$, and the prior for $\rho$ is $\log(\frac{1+\rho}{1-\rho}) \sim N(0, 0.15)$. The SPDE hyper-parameters are represented as $\log(\tau_\omega) = \theta_1$, and $\log(\kappa) = \theta_2$ in INLA, with $\theta_1, \theta_2 \sim N(0, 1)$.

\section{Simulation Study}\label{sec:simulation}

\subsection{Simulation Scenarios}

This section presents a simulation study designed to evaluate the performance of our methodology. The study serves two primary objectives: (1) to evaluate the accuracy of the parameter estimates, and (2) to assess the performance of predictions. It is crucial to ensure that the proposed model performs well not only in making accurate predictions on unobserved data, but also in 
parameter estimation. The simulation is designed to mimic the actual concentration levels of chlorophyll-a (chl-a), which are a key indicator of cyanobacteria abundance. 
In practice, measurements of chl-a may be directly collected at specific points near monitoring stations, or alternatively acquired from satellite pixels over larger geographical areas. We use Western Lake Erie as our study region, as this area is also the focus of our subsequent real data application.

\subsubsection{Simulation of the spatio-temporal GMRF}

First, we create an SPDE mesh on the spatial domain with a fine inner extension of 0.05 and outer extension of 0.2 to ensure the precision of the resulting field. Following the mesh creation, we simulate a zero-mean Gaussian Markov Random Field $\bm{\xi_ 1}$ on the SPDE mesh. The spatial correlations within this field are governed by the Mat\'ern covariance function, with parameters set to $\kappa = 7$ and $\sigma_\omega^2 = 0.25$. In addition, we adopt the default $\alpha = 2$ as the order of the differential operator in the SPDE. The configuration dictates a moderate level of smoothness and variability inherent in the spatial structure of the Gaussian field.
 
\begin{landscape}
\begin{figure}[h]
    \centering
    \begin{subfigure}[b]{1.35\textwidth}
         \centering
         \includegraphics[width=\textwidth]{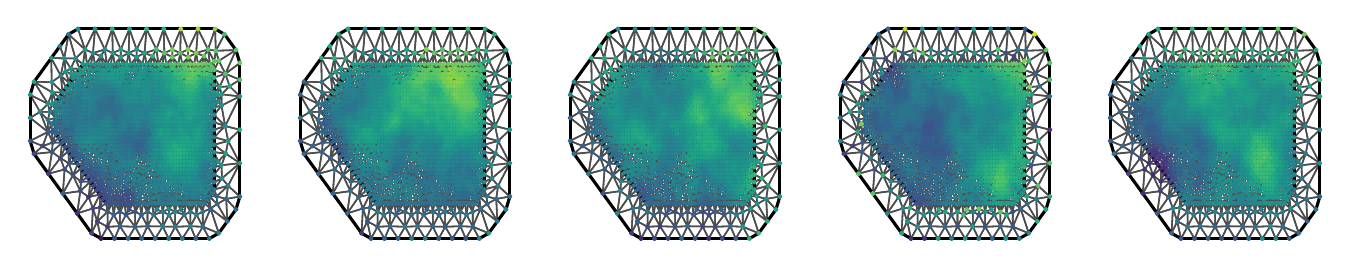}
         \caption{The latent spatio-temporal process $y(\bm{s}, t)$ on days 1, 5, 10, 15, and 19. }
         \label{fig:sim_plot_1}
     \end{subfigure}
     \begin{subfigure}[b]{1.35\textwidth}
         \centering
         \includegraphics[width=\textwidth]{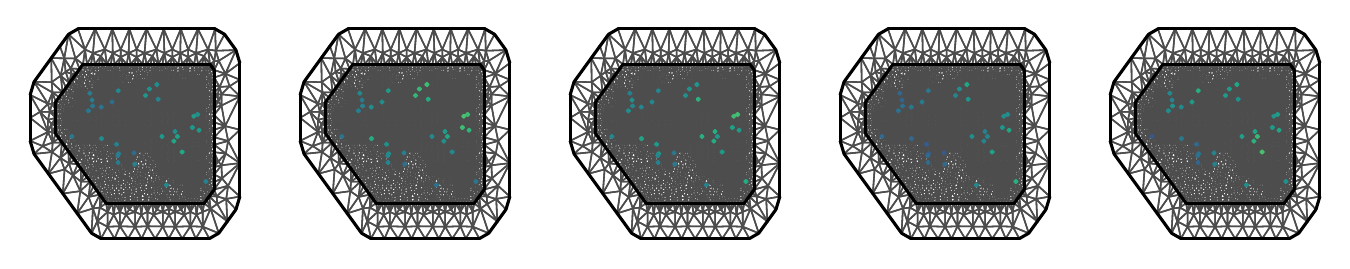}
         \caption{In situ samples on days 1, 5, 10, 15, and 19 (30 each day). The locations were selected randomly over the spatial domain.}
         \label{fig:sim_plot_2}
     \end{subfigure}
     \begin{subfigure}[b]{1.35\textwidth}
         \centering
         \includegraphics[width=\textwidth]{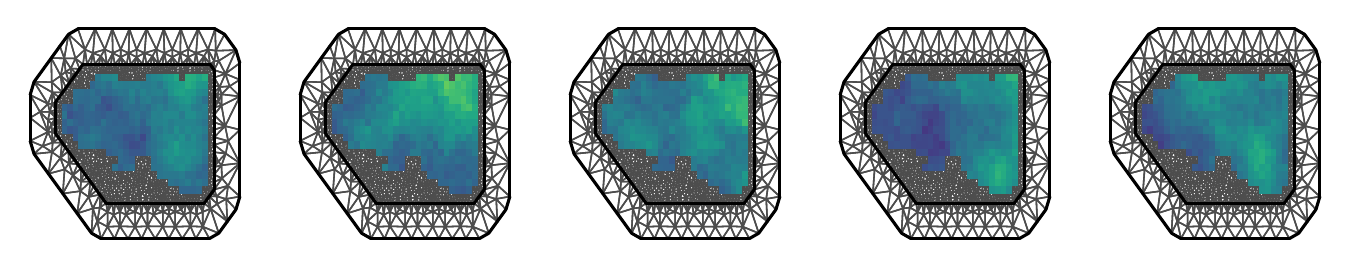}
         \caption{Satellite images with full observations on days 1, 5, 10, 15, and 19. Each pixel represents the average log chl-a level inside the block area.}
         \label{fig:sim_plot_3}
     \end{subfigure}
    \caption{A simulated realization of log chl-a concentration in Western Lake Erie on days 1, 5, 10, 15, and 19. The black lines represent the mesh triangulation. The colors superimposed on the mesh indicate the relative levels of log chl-a concentration. }
    \label{fig:sim_plot}
\end{figure}
\end{landscape}

We opt for a simulation duration of $T=19$ days, comprised of 14 days for training and 5 days for testing. This selection is made to strike a balance by ensuring a period of sufficient length for model training while maintaining a fast computation speed. 
Based on the selected $T$, we generate a realization of the spatio-temporal process sequentially according to \eqref{eqn:spatio-temporal}, with $\rho = 0.7$. Note that $\bm\omega_t = \sqrt{1-\rho}\bm{\xi_t}$ is also a GMRF with mean zero and variance $\sigma_\omega^2$, and this ensures the spatial process is stationary \citep{rue2005gaussian}. Moreover, to introduce a geographical component, we include the demeaned longitude and latitude as covariates;  
their regression coefficients are then set as $\beta_1 = -1$ and $\beta_2 = -1$, respectively. Consequently, the mean of the resulting simulated spatio-temporal process reflects the actual spatial trends characterized by an increase in chl-a concentration along the South and West of the basin. Figure \ref{fig:sim_plot_1} shows a realization of the simulated spatio-temporal GMRF on day 1, 5, 10, 15 and 19.

\subsubsection{Simulation of the in situ and satellite data}

We explore two settings for the sampling of in situ data: 5 vs.~30 sampling locations. The first scenario represents an extremely sparse sampling strategy. In the second scenario, while the sampling strategy is less sparse, it is designed to represent the upper limit of samples that might be collected in the field each day. This variation allows us to assess the impact of sampling intensity on the representation of the latent process at observed locations. The sampling locations could be treated as different or fixed over the $T$ days, and we adopt the latter as it is more likely to mimic the data generating mechanism of in situ samples. 
Once the locations are sampled, we interpolate the mesh nodes to align with the observation locations and subsequently generate in situ data according to \eqref{eqn:insitu}, with $\tau_2 = 50$ to introduce some small variability via the random noise term. Figure \ref{fig:sim_plot_2} shows an example of the simulated in situ data with 30 sampling locations.

The satellite data are generated based on the weighted average of the vertices of inside each rectangular grid in addition to a bias. We construct the mapping matrix $\bm{A^{a}}$ outlined in \eqref{eqn:mapping_a} and generate satellite data according to \eqref{eqn:satellite}, with $\tau_1 = 50$ and a constant bias $a = 0.5$ across space. We chose $\tau_y = \tau_1 = \tau_2$ for simplicity and to reduce the computation burden. Figure \ref{fig:sim_plot_3} shows an example of simulated satellite data, assuming all pixels are fully observed. It is important to note that a common challenge in satellite imagery is the presence of missing data, often caused by cloud cover. Consequently, we consider two settings for the simulation of satellite data with a different level of missingness per day: 50\% and 80\%. 

In addition, we consider three settings for the mesh, obtained by varying the lengths of the inner and outer extensions.  This variation is designed to assess the effect of mesh size on the parameter estimation and final prediction results. Since we are addressing COSP, the effect of mesh can be important as it affects how the weights are assigned among the triangulation vertices for approximating the integral of the areal data. Table \ref{tab:simulation_scenarios} summarizes the 12 simulation scenarios obtained as combinations of the three factors described above.

\begin{table}[h]
    \centering
    \begin{tabular*}{\textwidth}{c @{\extracolsep{\fill}} ccc}
    \hline
       Scenario & No. in situ samples & \% missingness & Mesh maximum inner length \\ \hline
        1 & 5 & 50\% & 0.15\\
        2 & 30 & 50\% & 0.15\\
        3 & 5 & 80\% & 0.15\\
        4 & 30 & 80\% & 0.15\\
        5 & 5 & 50\% & 0.1\\
        6 & 30 & 50\% & 0.1\\
        7 & 5 & 80\% & 0.1\\
        8 & 30 & 80\% & 0.1\\
        9 & 5 & 50\% & 0.05\\
        10 & 30 & 50\% & 0.05\\
        11 & 5 & 80\% & 0.05\\
        12 & 30 & 80\% & 0.05\\\hline
    \end{tabular*}
    \caption{Summary of the 12 simulation scenarios. The scenarios vary by three changing factors: the number of in situ samples collected each day, the percentage of missingness in satellite images, and the maximum inner length of the triangulation.}
    \label{tab:simulation_scenarios}
\end{table}

\subsection{Evaluation Metric}

The models were assessed on both parameter estimates and the predictive performance at unobserved locations. To evaluate the estimation performance for the model parameters, we measure the bias and the root mean squared error (RMSE) from the estimated posterior distribution of each parameter $\theta \in \{a, \bm\beta, \Theta\}$. We run $n_{sim}$ replications of the simulation, and denote each run (i.e., an independent realization of the spatio-temporal process) with the index $j$, where $j=1 \dots, n_{sim}$. For simulation run $j$, we denote each sample of $\theta$ as ${\hat{\theta}_{jk}}$, where $k=1, \dots, n_{samp}$. Each sample is drawn from the corresponding posterior distribution in the $j$-th simulation run. We chose $n_{sim}=100$ and $n_{samp}=100$ for the simulation study. Then the expressions for the bias and RMSE of a given parameter $\theta$ for each individual simulation run $j$ is:
\begin{equation}
    Bias_{\theta} = \frac{1}{n_{samp}}\sum_{k=1}^{n_{samp}}(\hat\theta_{jk} - \theta), \label{eq:par_bias}\\
\end{equation}
\begin{equation}
    RMSE_{\theta} = \sqrt{\frac{1}{n_{samp}}\sum_{k=1}^{n_{samp}}(\hat\theta_{jk} - \theta)^2}.  \label{eq:par_rmse}
\end{equation}

We also use RMSE to evaluate the predictive performance of the model. In each simulation run $j, j=1, \dots, n_{sim}$, denote $y_j(\bm{s}_l, t), \bm{s}_l \subset D, l=1,\dots, n_{pred}$ as the value of the simulated surface at location $\bm{s}_l$ and time $t$, and $\hat{y}_j(\bm{s}_l, t)$ is the prediction of $y_j(\bm{s}_l, t)$, i.e., the posterior mean from $p(\bm{y}(\bm{s}_l, t)|\bm{z}_1, \bm{z}_{2,obs})$ for $t \in \{1, \dots, 14\}$ and the posterior predictive mean from $p(\bm{y}(\bm{s}_l, t)|\bm{z}_1, \bm{z}_{2,obs})$ for $t \in \{15, \dots, 19\}$. The locations were randomly selected over the spatial domain $D$. We chose $n_{sim}=100$ and $n_{pred} = 20$ for the simulation study. For each simulation run $j$, the RMSE of the prediction at time $t$ over all $n_{pred}$ locations is calculated as follows:
\begin{equation}
    RMSE_{\bm{y}_t} = \sqrt{\frac{1}{n_{pred}}\sum_{l=1}^{n_{pred}}(\hat{y}_j(\bm{s}_l, t) - y_j(\bm{s}_l, t))^2}. \label{eq:pred_rmse}
\end{equation}

\subsection{Simulation Results}

Since the fusion model is proposed to reconcile spatial misalignment between in situ and satellite data, we assess its performance and effectiveness by comparing it with the standalone in situ and satellite models. This comparative analysis involves the estimation of parameters and prediction using each dataset in isolation. Therefore, we obtain the estimates and predictions for the standalone in situ model using \eqref{eqn:insitu}, \eqref{eqn:spatio-temporal-covariate}, \eqref{eqn:spatio-temporal} and for the standalone satellite model using \eqref{eqn:satellite}, \eqref{eqn:spatio-temporal-covariate}, \eqref{eqn:spatio-temporal}. Note that the bias parameter $a$ is not estimable in both standalone models. This is because in the in situ model \eqref{eqn:insitu}, $a$ is absent, and in the satellite model \eqref{eqn:satellite}, $a$ is subsumed into the parameter $\beta_0$. 

\subsubsection{Parameter estimation results}

\begin{figure}
    \centering
    \includegraphics[width=\textwidth]{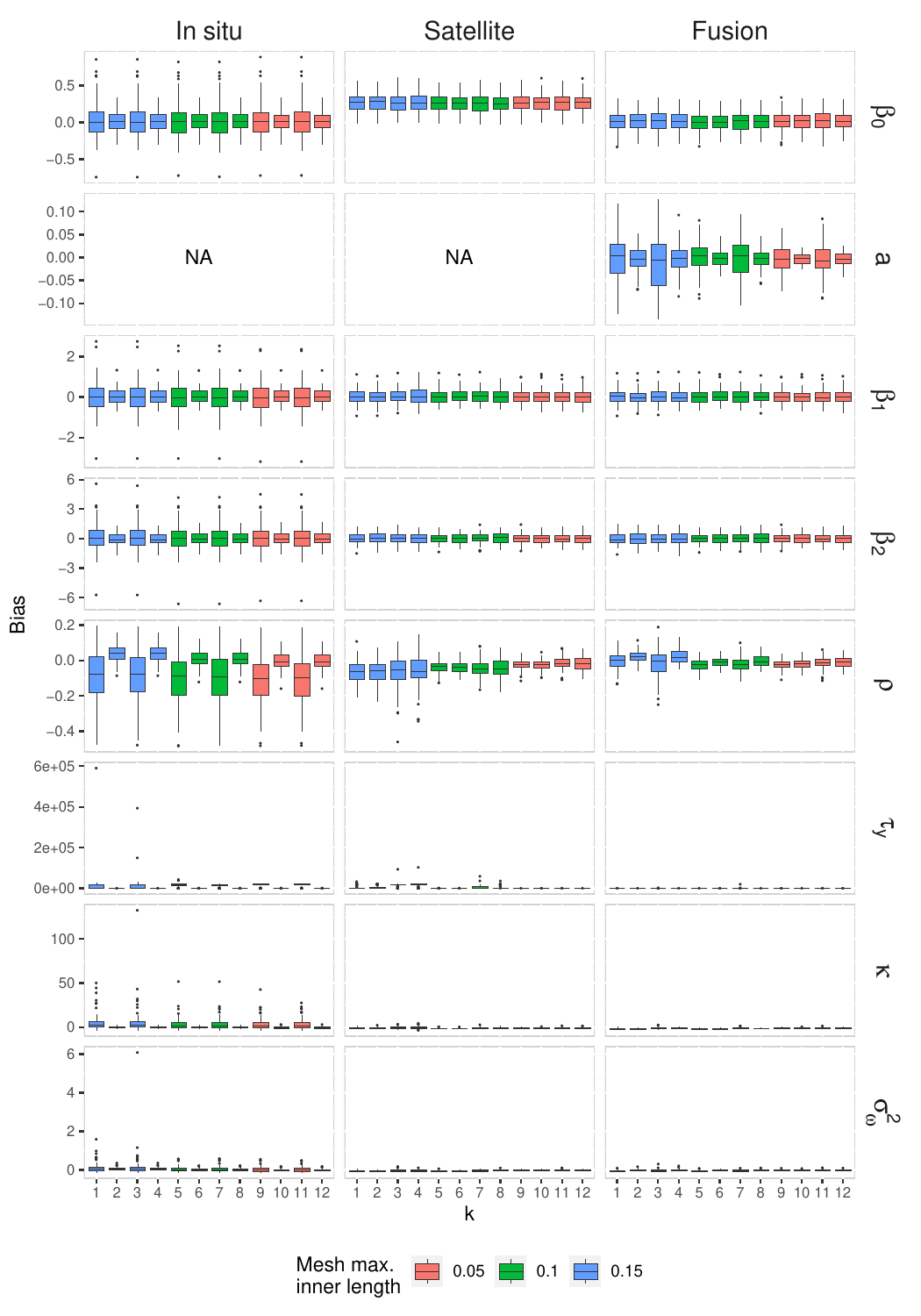}
    \caption{Boxplots for bias of parameters $\beta_0, a, \beta_1, \beta_2, \tau_z, \kappa, \sigma_\omega^2$ for the in situ, satellite, and fusion models. $k$ from 1 to 12 represents the simulation scenario number in Table \ref{tab:simulation_scenarios}. The boxplot colors represent the three different maximum inner lengths of the triangulation mesh.}
    \label{fig:param_bias}
\end{figure}

\begin{figure}
    \centering
    \includegraphics[width=\textwidth]{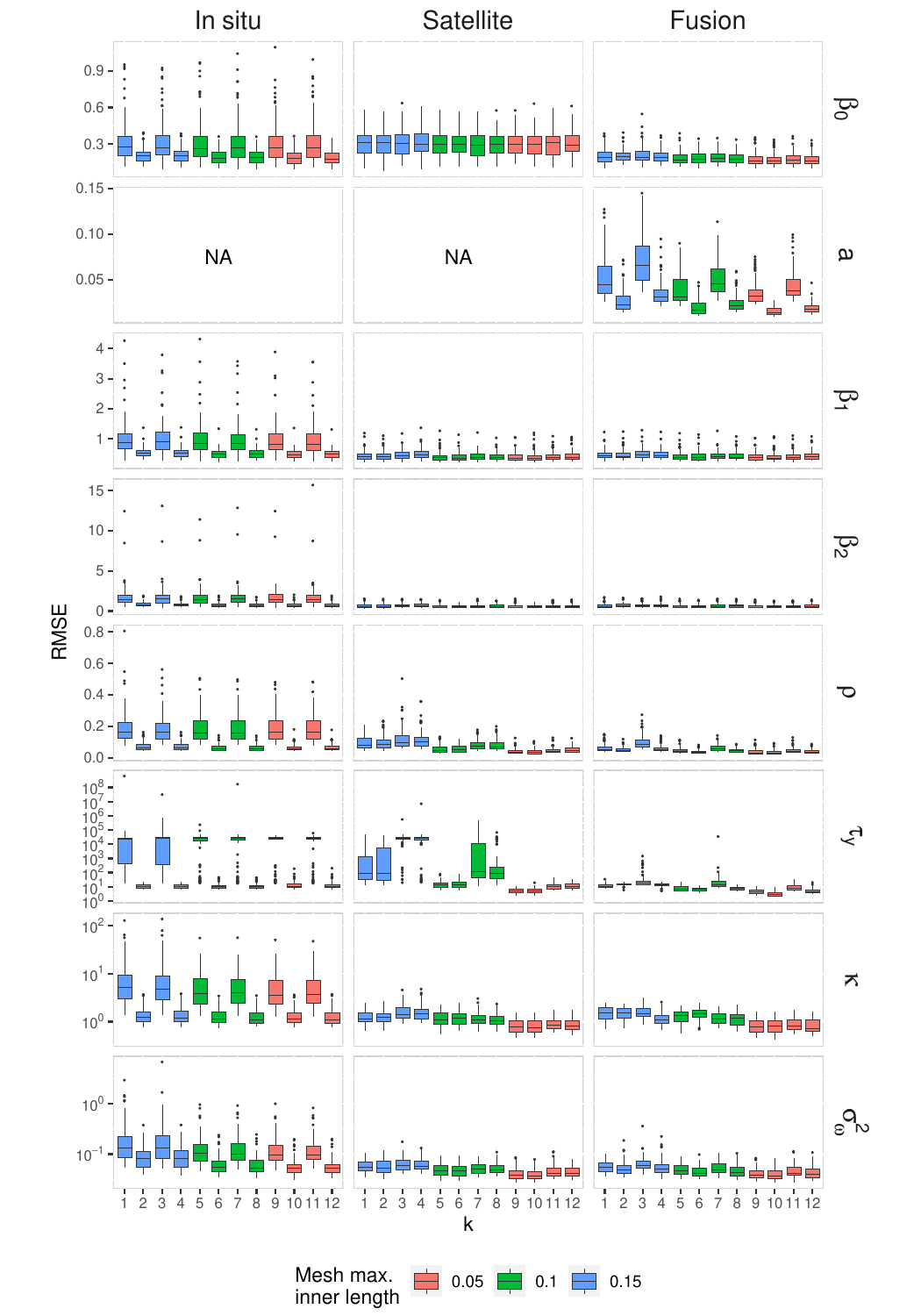}
    \caption{Boxplots for RMSE of parameters $\beta_0, a, \beta_1, \beta_2, \tau_z, \kappa, \sigma_\omega^2$ for the in situ, satellite, and fusion models. $k$ from 1 to 12 represents the simulation scenario number in Table \ref{tab:simulation_scenarios}. The boxplot colors represent the three different maximum inner lengths of the triangulation mesh. For visualization clarity, the parameters $\tau_y, \kappa, \sigma_\omega^2$ are displayed on a logarithmic scale.}
    \label{fig:param_rmse}
\end{figure}

Figures \ref{fig:param_bias} and \ref{fig:param_rmse} show boxplots of bias and RMSE of the parameter estimates for each individual simulation run calculated using \eqref{eq:par_bias} and \eqref{eq:par_rmse}. Our results indicate that all three simulation factors affect the bias and RMSE of each parameter. More specifically, both the parameter bias and RMSE for all three models tend to decrease as the mesh becomes more refined. The in situ model and fusion model display improved estimates with an increasing number of in situ samples, while the satellite and fusion model provide better estimates when the percentage of missingness decreases from 80\% to 50\%. 
While the missingness in satellite data is typically beyond control, the findings suggest the importance of choosing a smaller mesh size for computation (and if possible, increasing the number of in situ samples collected each day).

Further, it is evident from the boxplots that all three models show a relatively small bias and RMSE for the fixed effects $\beta_0, a, \beta_1, \beta_2$ and the hyper-parameter $\rho$. On the other hand, the bias and RMSE show a larger discrepancy across scenarios for the hyper-parameters $\tau_z, \kappa$ and $\sigma_\omega^2$, especially in the in situ model with only 5 observations, where extreme outliers tend to be present among the estimates. The results suggest $\beta_0, a, \beta_1, \beta_2$ and $\rho$ are less sensitive to changes in simulation settings, while $\tau_z, \kappa$ and $\sigma_\omega^2$ are more sensitive and should be interpreted with care when data is sparse.

Table \ref{tab:param_rmse} displays the average RMSEs of parameter estimates across the 100 simulation runs. In general, the fusion model provides better parameter estimates than both the in situ and the satellite models across the different scenarios. Notably, the fusion model exihibits the lowest RMSE for parameters $\beta_0, \tau_z, \rho, \sigma_\omega^2$ across all scenarios. However, for $\beta_1, \beta_2$ and $\kappa$, the fusion model may not consistently outperform the in situ or satellite models in certain scenarios. Particularly for $\kappa$, the fusion model may be inferior to the in situ model when the sample size is large and the mesh size is not refined.  For $\beta_1, \beta_2$ and $\kappa$, the fusion model outperforms the satellite model only in scenarios with 80\% missingness and the most refined mesh size. However, the real application most resembles scenario 11 and 12, where the RMSE is consistently smaller than the in situ and satellite models for all parameters.

Overall, the results indicate the importance of selecting a small mesh size when feasible. They also imply that, given the common occurrence of high missingness in satellite data, the fusion model is likely to yield the most accurate parameter estimates. 

\subsubsection{Prediction results}

\begin{figure}[ht]
    \centering
    \includegraphics[width=\textwidth]{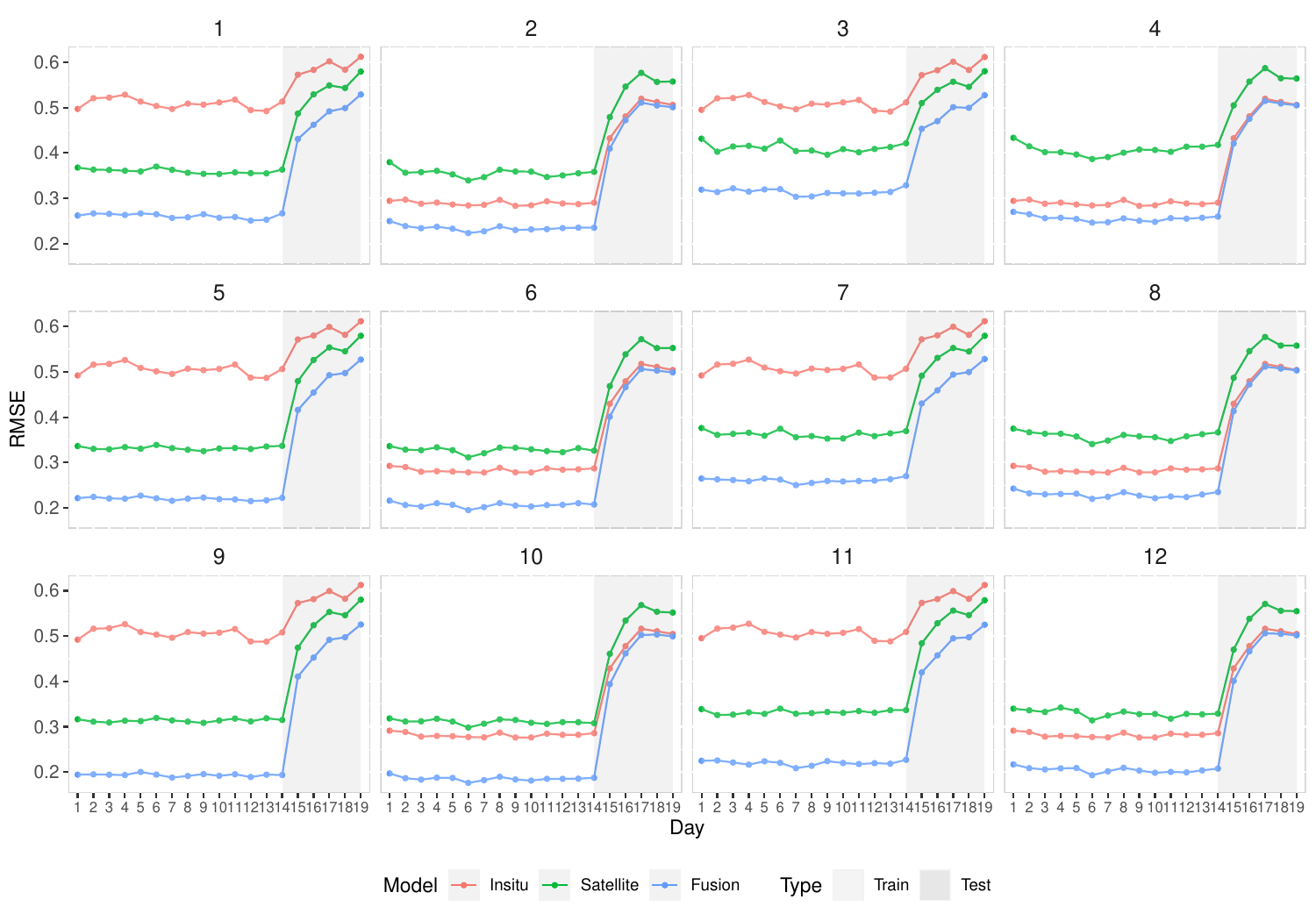}
    \caption{Prediction RMSE over unobserved locations on day 1-19 (with test days 15-19 shaded in grey; the training period is days 1-14). Panels 1-12 denote the different scenarios in Table \ref{tab:simulation_scenarios}. The lines colored in red, green, and blue correspond to the in situ, satellite, and fusion models, respectively.}
    \label{fig:pred_rmse_ts}
\end{figure}

Figure \ref{fig:pred_rmse_ts} shows the prediction results over unobserved locations on each day $t = 1, \ldots, 19$, averaged across the 100 simulation runs. Overall, the prediction accuracy from the fusion model outperforms those obtained from the in situ or satellite models in all scenarios. The smallest prediction RMSE occurs in scenario 10, i.e., when the number of in situ samples collected is the highest (30 per day), the percentage of missingness is the lowest (50\%), and the mesh size configuration is the smallest. 

Similar to parameter estimation, the size of the mesh configuration plays a role in the prediction RMSE. The smallest mesh configuration is observed to yield the lowest RMSE across the evaluated days (scenario 9 - 12), which indicates a small mesh is necessary to generate accurate prediction. Time-wise, the RMSE is expected to increase when transitioning from the training period (day 1 - day 14) to the test period (day 15 - day 19). This is because as we progress into the test period, predictions become less reliant on observations and more on noise terms as the correlation decays. Moreover, as we transit to the test period, the distinctions in RMSE between the different models and scenarios become less apparent, as illustrated in Figure \ref{fig:pred_rmse}.

\begin{figure}[h]
    \centering
    \includegraphics[width=.95\textwidth]{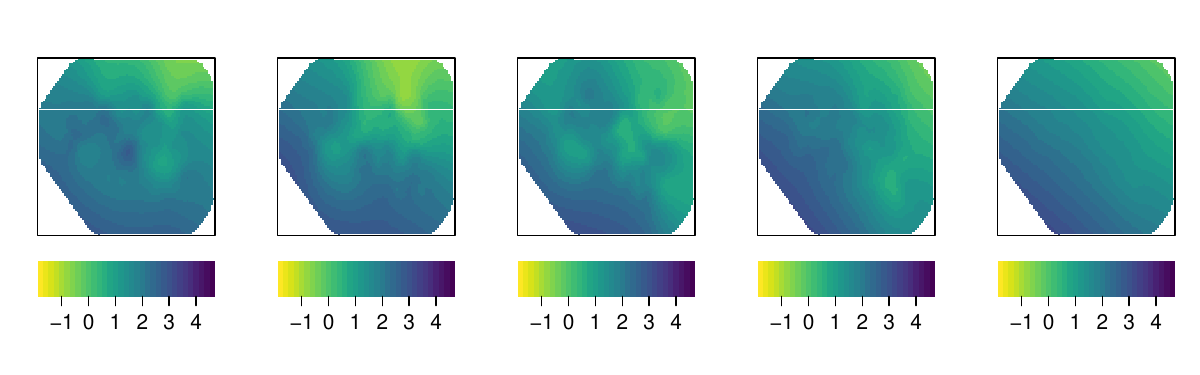}
    \includegraphics[width=.95\textwidth]{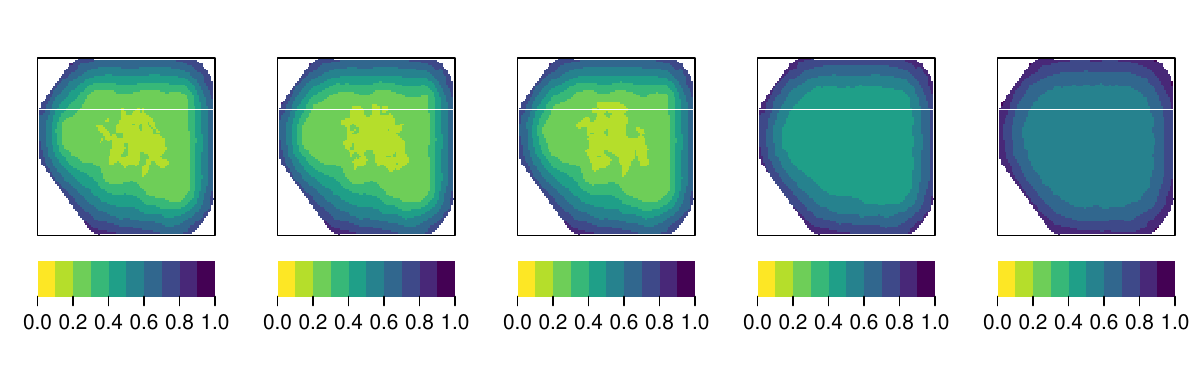}
    \caption{Posterior mean (top row) and standard deviation (bottom row) of the latent spatio-temporal field on day 1, 5, 10, 15 and 19 respectively. The color scale represents the posterior mean level (top row) and the standard deviation (bottom row) of the log chl-a concentration.}
    \label{fig:fitted}
\end{figure}

Figure \ref{fig:pred_rmse_ts} also suggests that predictions based on in situ data are not guaranteed to be more accurate than those based on satellite data, and vice versa. Utilizing only 5 in situ samples collected each day may not provide sufficient information for accurate estimation at unobserved locations, as the RMSE is observed to be highest for the in situ model when sampling is sparse (scenarios 1, 3, 5, 7, 9, and 11). However, as the sample size increases to 30, the prediction performance of in situ data has the potential to outperform satellite data, driven by the bias in the latter. The implications could be valuable for institutions heavily reliant on satellite data. Nevertheless, it is noteworthy that predictions based on the fusion model consistently demonstrate better performance, with a substantial margin on training days and a reduced margin on test days. This indicates that both in situ and satellite data are needed to generate more reliable predictions and draw accurate conclusions in such environmental studies.

\FloatBarrier
To visualize the results for an individual simulation run, Figure \ref{fig:fitted} shows the posterior mean and standard deviation for the estimated latent spatio-temporal field on day 1,5,10,15, and 19, under the fusion model fitted to the simulated data depicted in Figure \ref{fig:sim_plot}. The posterior mean panels are seen to closely reflect the original simulation output in Figure \ref{fig:sim_plot_1} and the increasing chl-a concentration observed along the Southern and Western basin. The posterior standard deviation panels reflect the increasing variability on test days 15-19, when compared with the training days 1-14. This emphasizes the fact that future predictions (i.e., during the test period) are more challenging and associated with greater uncertainty.

\FloatBarrier

\section{Application to the cyanobacteria HAB data}\label{sec:application}

We showcase an application of the method with a study focused on cyanobacteria HAB in Western Lake Erie. Lake Erie is one of the largest freshwater lakes in
the world, ranking 9$^\text{th}$ by area and 15$^\text{th}$ by volume \citep{herdendorf1990distribution,herdendorf1992lake}. It is the southernmost of the
North American Great Lakes, located between
41$\degree$21$^\prime$N and 42$\degree$50$^\prime$N latitude and 78$\degree$50$^\prime$W and
83$\degree$30$^\prime$W longitude. Based on bathymetry, Lake Erie is divided into western, central, and eastern basins. The western basin, lying west of a line from the tip of Pelee Point, Ontario, to Cedar Point, Ohio, is the smallest and the shallowest basin. 

Lake Erie has been observed to experience an increase in the frequency and intensity of cyanobacterial HAB events over the last decade \citep{bertani2017tracking, michalak2013record}. The most severe and intense lake-wide cyanobacteria bloom in Lake Erie occurred in 2011 and 2015, with a peak coverage of over 5000 km\textsuperscript{2} affecting both the U.S. and Canadian coastlines \citep{bertani2016probabilistically, stumpf2012interannual}. Moreover, a bloom in 2014 resulted in a loss of safe drinking water for almost 500,000 people living in the Toledo, OH area \citep{bullerjahn2016global}. To deepen our understanding of and mitigate such events, accurate estimation of the underlying drivers and precise predictions are crucial. Successfully achieving this objective hinges on the effective utilization and integration of both in situ and satellite data.

\begin{figure}[h]
    \centering
    \begin{subfigure}{.475\textwidth}
        \includegraphics[width=\textwidth]{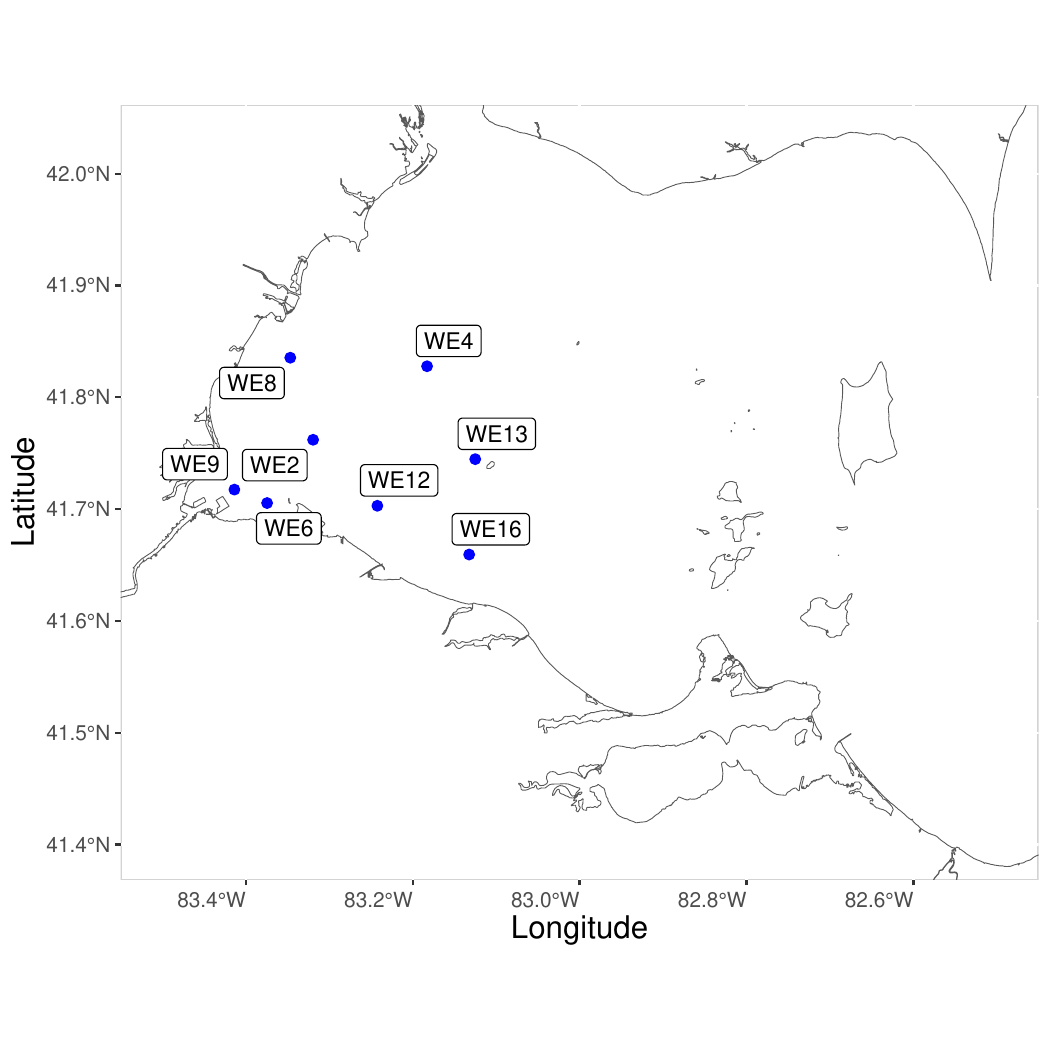}
        \caption{NOAA monitoring stations in Western Lake Erie.}
        \label{fig:insitu_site}
    \end{subfigure}
    \begin{subfigure}{.46\textwidth}
        \includegraphics[width=\textwidth]{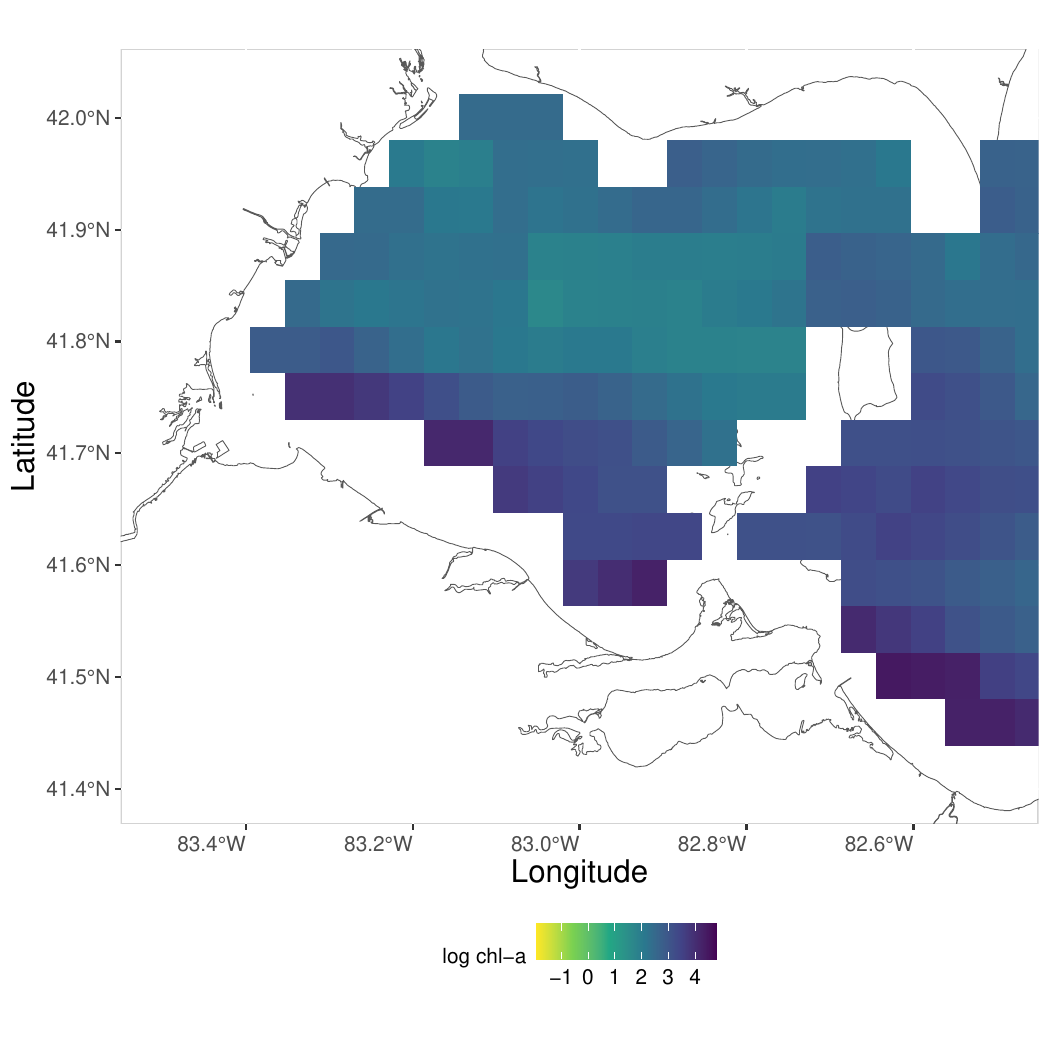}
        \caption{NASA MODIS level-3 chl-a products for Western Lake Erie on Aug 13th, 2015.}
        \label{fig:satellite_site}
    \end{subfigure}  
    \caption{In situ monitoring stations and a satellite image example for the study of cyanobacteria HABs in Western Lake Erie.}
    \label{fig:app}
\end{figure}

We obtained the in situ data from the in situ monitoring programs at NOAA (National Oceanic and Atmosphere Administration, Great Lakes Environmental Research Laboratory). Figure \ref{fig:insitu_site} indicates the locations of in situ samples collected over the bloom season. The satellite image data in our study were obtained from NASA Moderate-Resolution Imaging Spectroradiometer (MODIS) Level-3 chl-a products. They were aggregated and projected onto a spatial grid, with each pixel representing the average chl-a in a resolution of 4 km $\times$ 4 km in Lake Erie. Figure \ref{fig:satellite_site} shows an example of the satellite image over Western Lake Erie on Aug 13th, 2015. On average, these images are limited by 80\% cloud coverage, which suggests considerable uncertainties associated with remote sensing measurements.

We fit the combined datasets from May 2014 to May 2016 to the fusion model presented in this paper, incorporating relevant covariates that include longitude, latitude, basin-wide wind speed, air temperature, and surface temperatures as covariates. The findings are summarized in Table \ref{tab:app}. Our results indicate a notable misalignment between in situ and satellite observations, as the 95\% credible interval (CI) for the bias term associated with satellite images does not contain zero. This result necessitates the consideration of such a fusion model for an accurate analysis of harmful algae bloom events. We also find that the covariates longitude, latitude, and basin-wide surface temperatures are significant predictors for an increase in chl-a concentration.

\begin{table}[h]
    \centering 
    \begin{tabular*}{\textwidth}{l @{\extracolsep{\fill}} lccc}
    \hline
        Parameter & Mean & Std & 95\% CI \\ \hline
        $\beta_0$ (Intercept) & 1.06 & 0.11 & [0.85, 1.27] \\ 
        {$a$ (Bias for satellite images)} & {-0.14} & {0.01} & {[-0.17, -0.12]} \\
        {$\beta_1$ (Longitude)} & {-0.51} & {0.08} & {[-0.68, -0.35]} \\
        {$\beta_2$ (Latitude)} & {-1.30} & {0.12} & {[-1.53, -1.06]} \\
        $\beta_3$ (Wind Speed) & -0.01 & 0.01 & [-0.04, 0.01] \\
        $\beta_4$ (Air Temperature) & -0.01 & 0.01 & [-0.02, 0.01] \\ 
        {$\beta_5$ (Surface Temperature)} & {0.06} & {0.01} & {[0.04, 0.08]} \\
        $\tau_z$ & 166.87 & 3.83 & [159.42, 174.50] \\ 
        $\rho$ & 0.58 & 0.01 & [0.56, 0.60] \\
        $\kappa$ & 4.81 & 0.10 & [4.62, 5.00] \\
        $\sigma_\omega^2$ & 0.36 & 0.01 & [0.34, 0.39] \\ \hline
    \end{tabular*}
    \caption{Fusion model results for the parameter estimates obtained from the log chl-a concentration data during May 2014 - May 2016. The columns show the posterior mean, posterior standard deviation, and the 95\% CI for each parameter.}
    \label{tab:app}
\end{table}

\FloatBarrier

\section{Discussion}\label{sec:discussion}

In this work, we propose a data fusion method for in situ and satellite data on a spatio-temporal scale, which provides an extension to the work in \cite{moraga2017geostatistical}. Such an extension is necessary as most environmental data has both spatial and temporal dimensions, and relying solely on a snapshot of the spatial data often fails to provide a comprehensive understanding of the underlying dynamics. The fusion method assumes a latent spatio-temporal process that captures the spatial and temporal dependencies while also accommodating large-scale covariates. The model is fitted using the INLA-SPDE approach to allow for fast and scalable computation. The combined in situ and satellite data are mapped via projection matrices that link the satellite and in situ data to the triangulation nodes of the underlying GMRF. Since the INLA-SPDE approach reduces the computational burden compared to traditional MCMC methods, we can effectively handle large spatio-temporal datasets with dimensions up to $10^5 - 10^6$ for the latent Gaussian field $\bm{\mathcal{X}}$.

Through simulations, we show the fusion model tends to provide better parameter estimates and predictions. Both the parameter estimation and prediction accuracy were shown to be affected by the mesh size, the number of in situ samples, and the percentage of missingness in the satellite data. For parameter estimations, fixed effects and $\rho$ show less sensitivity to changes in simulation settings, while $\tau_z, \kappa$ and $\sigma_\omega^2$ are more sensitive and should be interpreted with care when data is sparse. The simulation study also demonstrates the fusion model generally outperforms the standalone in situ and satellite models on parameter estimation, particularly in scenarios resembling real-world applications. Regarding prediction accuracy, the simulation study shows that the fusion model outperforms the in situ and satellite models across all scenarios, and that the relative improvement is more pronounced during the training period than the test period. In general, the gain in prediction accuracy will likely depend on $\rho$ and the lead time of predictions. 

We apply the method to the address the COSP of chl-a concentration obtained from in situ and satellite data to better quantify the HAB events. By integrating data from NOAA monitoring stations and NASA chl-a level-3 products, we show strong evidence for a bias in satellite data, and that the inclusion of both datasets in the analysis of HAB events is necessary. Within this field, we believe the proposed fusion method can contribute to the construction of a more robust algae bloom index, going beyond the limitations of simplistic categorizations as ``in situ" or ``remote sensing". Furthermore, it has the potential to enhance the unified characterization of bloom dynamics and aid in identifying the key drivers of HAB events.

Future work can be extended to address the misalignment in time between different data sources. This requires the model to be set up in continuous time or on a set of time knots \citep{blangiardo2015spatial}, where satellite data -- often representing daily averages or spanning several hours -- could be formulated as $z_1(\bm{B}_j, t) = a(\bm{B}_j, t) + \frac{1}{|u|}\frac{1}{|\bm{B}_j|}\int_{u} \int_{\bm{B}_j} y (\bm{s}, u) d\bm{s}du + e_1(\bm{B}_j, t), |u|>0, t=1, \dots, T$. The consideration of misalignment is crucial when the measurements have high fluctuations during a day, as seen in variables like temperature. Other types of data can be considered in such a framework, such as non-stationary, non-isotropic and non-Gaussian data. These extensions are non-trivial, due to the computational complexities of large-scale spatio-temporal models. Generally speaking, methodologies for the integration of point-level and area-level referenced data remains a significant and partially unresolved challenge within the field of spatial statistics. 

\section*{Acknowledgements}

We thank Serghei Bocaniov for helpful discussions and providing us with relevant algae bloom data.
This work was partially supported by the Natural Sciences and Engineering Research Council of Canada under Discovery Grant RGPIN-2019-04771.

\bibliographystyle{elsarticle-harv}
\bibliography{ref.bib}

\newpage
\appendix
\section{Simulation results}

\begin{figure}[ht]
    \centering
    \includegraphics[width=\textwidth]{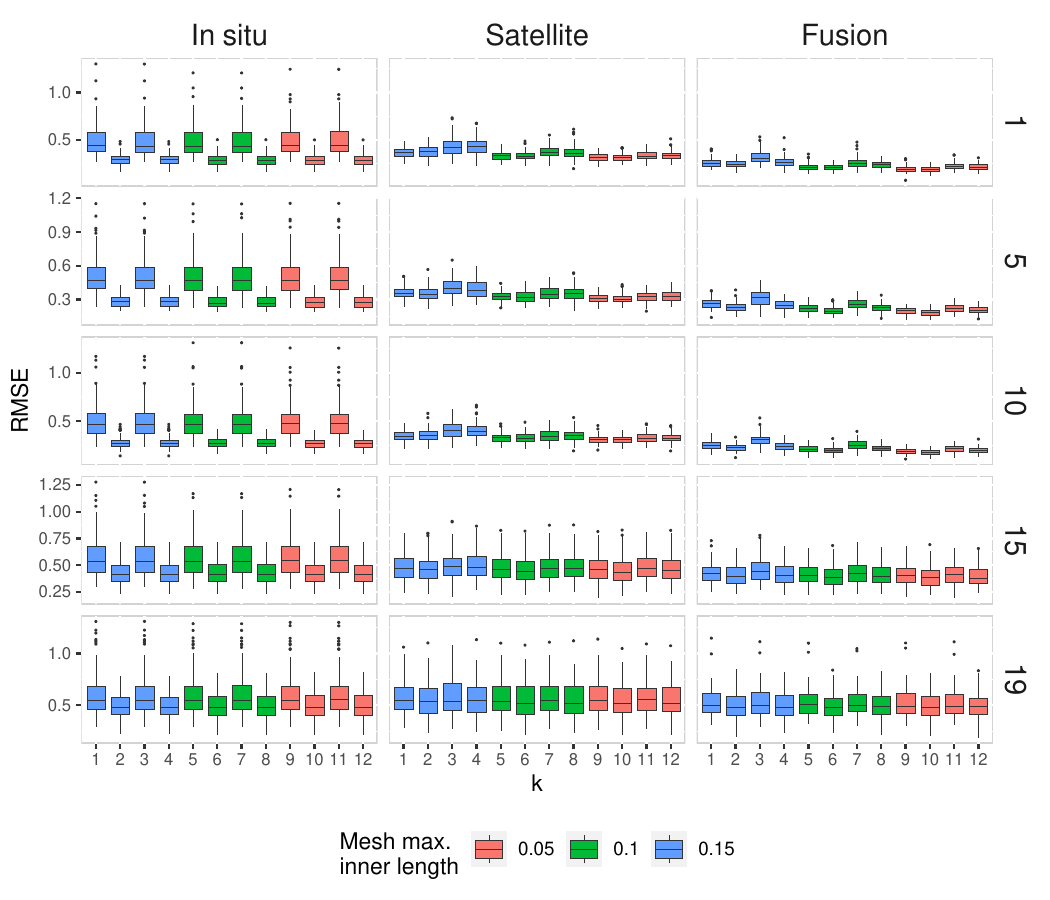}
    \caption{Prediction RMSE (calculated using \eqref{eq:pred_rmse}) on day 1, 5, 10, 15, and 19. $k$ from 1 to 12 represents the simulation scenario number in Table \ref{tab:simulation_scenarios}. The color in boxplot represents the three different maximum inner length of the triangulation. }
    \label{fig:pred_rmse}
\end{figure}

\begin{table}[ht]
\scriptsize
\centering
\begin{tabular}{lllll}
  \hline
Parameter & k & In situ & Satellite & Fusion \\ 
  \hline
  $\beta_0$ & 1 & 0.3237 & 0.3072 & 0.2018 \\ 
  $\beta_0$ & 2 & 0.2098 & 0.3085 & 0.206 \\ 
  $\beta_0$ & 3 & 0.3259 & 0.3123 & 0.2155 \\ 
  $\beta_0$ & 4 & 0.2097 & 0.3104 & 0.1997 \\ 
  $\beta_0$ & 5 & 0.3169 & 0.3005 & 0.1848 \\ 
  $\beta_0$ & 6 & 0.1963 & 0.2997 & 0.1887 \\ 
  $\beta_0$ & 7 & 0.3153 & 0.2994 & 0.19 \\ 
  $\beta_0$ & 8 & 0.198 & 0.3007 & 0.1834 \\ 
  $\beta_0$ & 9 & 0.3119 & 0.3031 & 0.175 \\ 
  $\beta_0$ & 10 & 0.1913 & 0.3022 & 0.1728 \\ 
  $\beta_0$ & 11 & 0.3135 & 0.3066 & 0.1796 \\ 
  $\beta_0$ & 12 & 0.1907 & 0.3046 & 0.1749 \\ 
  $a$ & 1 & NA & NA & 0.0553 \\ 
  $a$ & 2 & NA & NA & 0.0273 \\ 
  $a$ & 3 & NA & NA & 0.071 \\ 
  $a$ & 4 & NA & NA & 0.0359 \\ 
  $a$ & 5 & NA & NA & 0.0402 \\ 
  $a$ & 6 & NA & NA & 0.0204 \\ 
  $a$ & 7 & NA & NA & 0.0515 \\ 
  $a$ & 8 & NA & NA & 0.0244 \\ 
  $a$ & 9 & NA & NA & 0.0361 \\ 
  $a$ & 10 & NA & NA & 0.0159 \\ 
  $a$ & 11 & NA & NA & 0.045 \\ 
  $a$ & 12 & NA & NA & 0.0191 \\ 
  $\beta_1$ & 1 & 1.0464 & 0.4521 & 0.4816 \\ 
  $\beta_1$ & 2 & 0.5531 & 0.4564 & 0.4879 \\ 
  $\beta_1$ & 3 & 1.0282 & 0.4858 & 0.5124 \\ 
  $\beta_1$ & 4 & 0.5585 & 0.5186 & 0.4997 \\ 
  $\beta_1$ & 5 & 1.0174 & 0.4294 & 0.4363 \\ 
  $\beta_1$ & 6 & 0.5207 & 0.4208 & 0.4359 \\ 
  $\beta_1$ & 7 & 1.0097 & 0.4385 & 0.46 \\ 
  $\beta_1$ & 8 & 0.537 & 0.4372 & 0.4535 \\ 
  $\beta_1$ & 9 & 1.0058 & 0.4207 & 0.422 \\ 
  $\beta_1$ & 10 & 0.514 & 0.4206 & 0.419 \\ 
  $\beta_1$ & 11 & 1.001 & 0.4353 & 0.4342 \\ 
  $\beta_1$ & 12 & 0.5173 & 0.4414 & 0.4404 \\ 
  $\beta_2$ & 1 & 1.8043 & 0.6555 & 0.692 \\ 
  $\beta_2$ & 2 & 0.8427 & 0.6785 & 0.7343 \\ 
  $\beta_2$ & 3 & 1.8264 & 0.7365 & 0.7534 \\ 
  $\beta_2$ & 4 & 0.8371 & 0.7387 & 0.744 \\ 
  $\beta_2$ & 5 & 1.7831 & 0.5919 & 0.6065 \\ 
  $\beta_2$ & 6 & 0.8054 & 0.6055 & 0.6283 \\ 
  $\beta_2$ & 7 & 1.7965 & 0.6214 & 0.6413 \\ 
  $\beta_2$ & 8 & 0.7991 & 0.6596 & 0.6759 \\ 
  $\beta_2$ & 9 & 1.7803 & 0.5881 & 0.5931 \\ 
  $\beta_2$ & 10 & 0.7934 & 0.5781 & 0.5854 \\ 
  $\beta_2$ & 11 & 1.7883 & 0.6187 & 0.6087 \\ 
  $\beta_2$ & 12 & 0.7836 & 0.6312 & 0.6268 \\\hline
  \end{tabular}
  \begin{tabular}{lllll}
  \hline
  Parameter & k & In situ & Satellite & Fusion \\ 
  \hline 
  $\rho$ & 1 & 0.1959 & 0.0981 & 0.0659 \\ 
  $\rho$ & 2 & 0.0739 & 0.0992 & 0.0529 \\ 
  $\rho$ & 3 & 0.1895 & 0.1204 & 0.0998 \\ 
  $\rho$ & 4 & 0.0748 & 0.1187 & 0.0606 \\ 
  $\rho$ & 5 & 0.1879 & 0.0572 & 0.0488 \\ 
  $\rho$ & 6 & 0.0648 & 0.0599 & 0.0398 \\ 
  $\rho$ & 7 & 0.1889 & 0.0841 & 0.067 \\ 
  $\rho$ & 8 & 0.0649 & 0.0835 & 0.0494 \\ 
  $\rho$ & 9 & 0.189 & 0.0412 & 0.0407 \\ 
  $\rho$ & 10 & 0.0658 & 0.0425 & 0.0373 \\ 
  $\rho$ & 11 & 0.1898 & 0.05 & 0.0481 \\ 
  $\rho$ & 12 & 0.0659 & 0.0536 & 0.0437 \\ 
  $\tau_y$ & 1 & 6320326.3325 & 6131.3782 & 12.5873 \\ 
  $\tau_y$ & 2 & 11.9433 & 6745.899 & 16.0838 \\ 
  $\tau_y$ & 3 & 341727.665 & 31385.2875 & 61.0997 \\ 
  $\tau_y$ & 4 & 11.9916 & 95950.483 & 14.8899 \\ 
  $\tau_y$ & 5 & 25104.0128 & 17.3873 & 8.5355 \\ 
  $\tau_y$ & 6 & 12.3943 & 16.823 & 7.0195 \\ 
  $\tau_y$ & 7 & 1740424.0623 & 12571.966 & 374.0133 \\ 
  $\tau_y$ & 8 & 12.4525 & 3661.8357 & 8.3012 \\ 
  $\tau_y$ & 9 & 24183.8101 & 5.9555 & 5.4082 \\ 
  $\tau_y$ & 10 & 16.413 & 5.7937 & 3.3512 \\ 
  $\tau_y$ & 11 & 25230.9393 & 12.7219 & 10.4918 \\ 
  $\tau_y$ & 12 & 16.4875 & 13.9742 & 5.748 \\ 
  $\kappa$ & 1 & 9.7076 & 1.2978 & 1.5753 \\ 
  $\kappa$ & 2 & 1.4546 & 1.2992 & 1.5278 \\ 
  $\kappa$ & 3 & 9.8864 & 1.6252 & 1.6489 \\ 
  $\kappa$ & 4 & 1.446 & 1.6556 & 1.219 \\ 
  $\kappa$ & 5 & 6.4121 & 1.2099 & 1.3378 \\ 
  $\kappa$ & 6 & 1.3711 & 1.2264 & 1.494 \\ 
  $\kappa$ & 7 & 6.461 & 1.1941 & 1.2373 \\ 
  $\kappa$ & 8 & 1.3545 & 1.1541 & 1.2071 \\ 
  $\kappa$ & 9 & 6.3736 & 0.8616 & 0.8589 \\ 
  $\kappa$ & 10 & 1.3337 & 0.8397 & 0.867 \\ 
  $\kappa$ & 11 & 6.6266 & 0.9339 & 0.91 \\ 
  $\kappa$ & 12 & 1.3178 & 0.9059 & 0.8615 \\ 
  $\sigma_\omega^2$ & 1 & 0.2502 & 0.0605 & 0.06 \\ 
  $\sigma_\omega^2$ & 2 & 0.0933 & 0.0592 & 0.0546 \\ 
  $\sigma_\omega^2$ & 3 & 0.2814 & 0.0664 & 0.068 \\ 
  $\sigma_\omega^2$ & 4 & 0.0938 & 0.0658 & 0.0603 \\ 
  $\sigma_\omega^2$ & 5 & 0.1476 & 0.0512 & 0.0506 \\ 
  $\sigma_\omega^2$ & 6 & 0.066 & 0.0508 & 0.0466 \\ 
  $\sigma_\omega^2$ & 7 & 0.149 & 0.0533 & 0.0551 \\ 
  $\sigma_\omega^2$ & 8 & 0.0663 & 0.0541 & 0.0485 \\ 
  $\sigma_\omega^2$ & 9 & 0.1367 & 0.042 & 0.0424 \\ 
  $\sigma_\omega^2$ & 10 & 0.0603 & 0.0421 & 0.042 \\ 
  $\sigma_\omega^2$ & 11 & 0.1383 & 0.0483 & 0.0485 \\ 
  $\sigma_\omega^2$ & 12 & 0.0604 & 0.0466 & 0.0453 \\ 
   \hline
\end{tabular}
\caption{Average parameter RMSE across 100 simulation runs for the in situ, satellite, and fusion models. $k$ from 1 to 12 represents the simulation scenario number in Table \ref{tab:simulation_scenarios}.}
\label{tab:param_rmse}
\end{table}

\end{document}